\documentclass[aps,pre,onecolumn,showpacs,floatfix]{revtex4}
\usepackage{graphicx,amsmath,amssymb,mathptm,mathrsfs}
\begin{document}
\title{The finite density scaling laws of condensation phase transition in zero range processes on scale-free networks}
\author{Guifeng Su$^{1}$}
\author{Xiaowen Li$^{1}$}
\author{Yi Zhang$^{1}$}
\email{yizhang@shnu.edu.cn}
\author{Xiaobing Zhang$^{2}$}
\affiliation{${1}$ Department of Physics, Shanghai Normal University, Shanghai 200234, P. R. China \\
             ${2}$ School of Physics, Nankai University, Tianjin 300071, P. R. China}

%\date{\today}

\begin{abstract}
The dynamics of zero-range processes on complex networks is expected to be influenced by the topological structure
of underlying networks. A real space complete condensation phase transition in the stationary state may occur.
We have studied the finite density effects of the condensation transition in both the stationary and dynamical
zero-range process on scale-free networks. By means of grand canonical ensemble method, we predict analytically
the scaling laws of the average occupation number with respect to the finite density for the steady state. We
further explore the relaxation dynamics of the condensation phase transition. By applying the hierarchical evolution
and scaling ansatz, a scaling law for the relaxation dynamics is predicted. Monte Carlo simulations are performed
and the predicted density scaling laws are nicely validated.
\end{abstract}

\pacs{89.75.Hc, 05.20.-y, 02.50.Ey}
% 89.75.Hc : Networks and genealogical trees
% 05.20.-y : Classical statistical mechanics
% 02.50.Ey : Stochastic processes
\maketitle

\section{Introduction}

Condensation phase transitions are abundant phenomena in nature and observed in various physical circumstances
and stochastic processes. For instance, the Bose-Einstein condensation (BEC) in cold dilute atomic
gases~\cite{BEC}, jamming in traffic flow~\cite{Chow00} and granular flow~\cite{Meer02,Meer04}, condensation
of links in complex networks~\cite{Krap00,Bian01,Zhang12}, to name a few. Although the equilibrium condensation
phase transitions can be well described in the framework of equilibrium statistical mechanics~\cite{Path11},
building a consistent theory of nonequilibrium (condensation) phase transitions remains a big challenge.
How such transitions arise and what dynamical nature they may have are particularly intriguing questions to be
answered.
The long lasting efforts have been paid on the investigation of interacting particle systems~\cite{Ligg99}.
One of the important categories is so-called driven diffusive systems~\cite{Schm95}. These systems are driven
away from equilibrium by external forces, and may evolve toward a nonequilibrium steady state. In addition,
they exhibit non-trivial behavior such like phase separation~\cite{Evans98} or phase transitions (even in one
spatial dimension)~\cite{Evans06}.

Among above mentioned systems, the zero range processes (ZRPs), introduced by F. Spitzer~\cite{Spitz70}
roughly half century ago, has attracted considerable attention due to the fact that it is one of few rare
examples of analytically tractable nonequilibrium models. For recent reviews, see e.g.,~\cite{Evans05,Godre07}
and references therein.
%It should be noticed as well that ZRPs are defined by the dynamics, instead of Hamiltonian as in general
%equilibrium statistical mechanics. The stationary state of a ZRP is hence not specified by any quantum
%mechanical or classical distributions.
The prototype of ZRP is a stochastic particle system defined on one dimensional regular lattice of given
number of sites and particles hop onto the nearest neighbours (NNs) with some transition rate, say, $u(n)$.
This rate function depends only on the occupation number $n$ of its current position, not that of the other
sites -- this defines the zero-range property of the process. If $u(n)$ is decreasing in $n$, then an
effective attraction between particles is presented. As a result, a condensation transition occurs if $\rho$
exceeds some critical density $\rho_c$, and a macroscopically finite fraction of particles condense on a
single site. The condensation phase transitions in ZRPs have been studied based on periodic lattices
in~\cite{Evans00,Gross03,Majum05,Godre12}.

From the perspective of complex networks, the underlying periodic lattice of ZRPs is simply regular fully
connected network.
%when the condensation transition occurs in the stationary state of ZRP,
%its dynamical properties and the long-time evolution are still less understood.
%The condensation phase is exhibited in the form of a bump occurring near the tail of the distribution
%function~\cite{Evans05}. Most analysis of condensation in ZRPs have been conducted based on the grand
%canonical ensemble (GCE) in which the thermodynamic limit is already taken into consideration~\cite{Evans00,Evans05}.
However, many natural and manmade networks are self-organized as the scale-free (SF) networks (see e.g.,
\cite{BA02,Eguil05,Bocca06} and references therein).
%such as the Internet, airline transport system, and the world wide web,
The SF networks are strongly inhomogeneous and highly clustering in architecture~\cite{BA99}. They also
typically possess a power-law degree distribution, $\mathscr{P}_{deg}(k) \sim k^{-\gamma}$, where $k$ is
the degree (i.e. number of links) of a node and the constant $\gamma$ is degree exponent.
%critical phenomena in sf networks.
It has been demonstrated that these significant features of SF networks are important factors affected a
multitude of critical phenomena~\cite{Golt08}, and as well as dynamical
processes~\cite{Barth04,Sood05,Noh05a,Noh05b,Tang06} defined on them. It was found when the power exponent
$\delta$ in the hopping rate $u(n)=n^{\delta}$ is smaller than a critical value $\delta_c$, that is, when
$\delta<\delta_c=1/(\gamma-1)$, %where $\gamma$ is the degree exponent of the network,
a finite fraction of particles are condensed into the node with the maximum degree, or the
\emph{hub}~\cite{Noh05a}.
%When $\delta = 0$, the fraction can even reach $1$, this is so-called complete condensation.
Such condensation phase transitions are driven by the interplay of the (effective) interaction between
particles and the topological structure disorder of the underlying SF networks. However, to the author'
knowledge, so far the analysis of the possible effects of the finite density are still absent. In fact,
one naturally expects the density plays an important role in condensation transition of ZRPs on SF networks.
%Higher degree nodes usually have large values of the load or
%betweenness centrality, which is a measure of a traffic along shortest
%pathways~\cite{Newman01,Goh02}. Whether under stochastic or deterministic dynamics,
%particles tend to concentrate on high degree nodes. Consequently, one expects that a
%particle interaction would play an important role in dynamic processes. However, only
%interacting dynamical systems on SF networks~\cite{Holme03}.

In current paper we then focus on the finite density aspect of the condensation transition of the static
and dynamical ZRP defined on SF networks.
% we have to remind the readers that they should not get confused with the \emph{finite size} effects.
% skip ...
As we will show in the following, based on the grand canonical
ensemble (GCE) approach, we predict a novel scaling law in the condensation phase transition with
respect to (w.r.t.) the finite density.
%and the density dependent scaling relations for the relaxation dynamics are predicted as well.
Another important issue we addressed in this paper is the relaxation dynamics in the condensed phase
transition. A hierarchical evolution occurs in relaxation dynamics, i.e., the relaxation proceeds from
small degree nodes to larger and larger degree nodes. We predict that there exist the scaling laws of
the average occupation number and the inverse participation ratio (IPR) w.r.t. the finite density in
both steady state and relaxation dynamics.
To verify our analytical predictions, Monte Carlo simulations are fulfilled for steady state and the
relaxation dynamics, respectively. These predicted finite density scaling laws are nicely confirmed.

The rest of the paper is organized as follows: in Sec. II, we present the ZRP model on SF network, and the
corresponding GCE theoretical formalism. In Sec. III, we first argue the existence of the condensation phase
transition, and then predicted the finite density scaling laws for such transition. Monte Carlo simulations
are followed and fully confirmed previous analytical results. In Sec. IV, we explore the relaxation dynamics
both analytically and numerically. Predictions on the finite density scaling law for relaxation dynamics
are presented and simulated.
Finally in Sec.~\ref{sec:5}, we conclude the paper with the summary of our findings and some remarks.

\section{ZRP on SF Network}

\subsection{The Model}

Let us now consider the ZRP with $N=\rho L$ particles hopping on an SF network, with $\rho$ being the density
of the particles, and $L$ the number of nodes of the underlying SF networks. One has to keep in mind that in
current situation, the density $\rho$ is not fixed but variable.
The degree (probability) distribution of the SF networks is characterized by power-law distribution,
\begin{equation}
\label{eq:Pk}
\mathscr{P}_{deg} (k) = c k^{-\gamma} \,, ~k_\mathrm{min} \le k \le k_\mathrm{max} ,
\end{equation}
where $c$ is some appropriate normalization constant. For any finite network, the minimum degree $k_\mathrm{min}$
sets up the smallest possible degree and is a constant of order one, $k_\mathrm{min} \sim \mathscr{O}(1)$.
Recall that the node with the maximum degree is called hub, and its degree $k_\mathrm{max}$ satisfies~\cite{Cohen00}
\begin{equation}
\label{eq:kmax}
\int_{k_{max}}^{\infty} \mathscr{P}_{deg}(k) ~dk = \frac{1}{L} \,,
\end{equation}
such that $k_\mathrm{max} \approx k_{min} L^{1/(\gamma-1)}$. In general, one has $1 \leq k_{min} \ll k_{max}$.
We use $w_i$ to represent the weight of the $i$-th node of the network and $w_i=k_i/\sum_i k_i \equiv k_i/\sum k$,
where a shorthand notation $\sum k$ is used for convenience (the same below).
The average degree $\overline{k}$ of SF networks is defined as
\begin{equation}
\label{eq:kave}
\overline{k} \equiv \int dk~k~\mathscr{P}_{deg}(k) \,.
\end{equation}
To keep $\overline{k}$ finite, the degree exponent $\gamma >2$ is required.

One of the important properties of the ZRP stationary state probability $\mathscr{P}(\{n_i\})$, i.e., the
probability of finding particles in a microscopic configuration $\{n_i\}$, is that $\mathscr{P}(\{n_i\})$
takes the factorized form~\cite{Evans04,Zia04}, i.e.,
\begin{equation}
\label{eq:Pn}
\mathscr{P}(\{n_i\}) = \frac{1}{Z_{L,N}} \prod_{i=1}^L f(n_i) \, ,
\end{equation}
in which the microscopic configuration $\{n_i\}=$ \{$n_1$, $n_2$, $\cdots$, $n_L$\}, $n_i$ is the number of
particles on the $i$-th node of the network. The normalization factor $Z_{L,N}$ plays the role of partition
function, and can be computed by summing the product $\prod_{i=1}^L f(n_i)$ over all microscopic
configurations $\{n_i\}$,
\begin{equation}
\label{eq:ZLN}
Z_{L,N} = \sum_{\{n_i\}}~\prod_{i=1}^L f(n_i) \delta\Bigg(\sum_{i=1}^{L} n_i - N\Bigg) \,.
\end{equation}
The $\delta$ function introduced in Eq.(\ref{eq:ZLN}) guarantees the conservation of total number of particles
by the dynamics, i.e., $\sum n = N$.
The factors appeared in the product form $f(n_j)$'s are the single-site (the $j$-th node) statistical weights,
and can be expressed in terms of hopping rate functions $u(n_j)$,
%\begin{equation}
%\label{eq:fn1}
%f(n_0)=1, \,=\prod_{j=1}^n \left[ \frac{W_j}{u(n_j)}\right]\, ,
%\end{equation}
\begin{equation}
\label{eq:fn1}
f(n_j) = \left\{
         \begin{array}{rl}
          \prod_{j=1}^n \left[\frac{W_j}{u(j)}\right] \,,  & \text{if} ~~j \geq 1 \,, \\
          1 \,,                                            & \text{if} ~~j = 0 \,.
\end{array} \right.
\end{equation}
where $W_j$ is the single particle stationary state probability distribution, associated with the jumping
probability $T_{i \rightarrow j}$. It can be written as
\begin{equation}
T_{i \rightarrow j} = \left\{
\begin{array}{rl}
1/k_i \,, & \text{if i and j are connected} , \\
0 \,,     & \text{otherwise} .
\end{array} \right.
\end{equation}
The subscript $i \rightarrow j$ in jumping probability means that particles hop from the $i$-th (departure) node
to the $j$-th (arrival) node. The detailed balance equations fully determine the $W_j$'s:
\begin{equation}
\label{eq:WT}
\sum_{i}~W_i T_{i \rightarrow j}  = \sum_{j} W_j T_{j \rightarrow i} \,\, , i=1, 2, \cdots, L \,,
\end{equation}
i.e., the outgoing probability flux exactly matches the incoming probability flux. The solution to Eq.(\ref{eq:WT})
is $W_i=k_i/\sum k=w_i$~\cite{Noh04}, i.e., the stationary state probability distribution just equals to the weight
of the node: the more weights a node has, the more often it will be visited by a random hopper.
For the ZRP on the SF networks, the hopping rate function of particles out of the $j$-th node takes the form
\begin{equation}
\label{eq:un}
u(n_j)=n_j^\delta \,,
\end{equation}
where $n_j$ represents the number of particles located in the current ($j$-th) node, and $\delta$ is a global
parameter controlling the interaction strength between particles. For particles with (effective) attractive
interaction, $u(n)$ grows sub-linearly w.r.t the number of particles $n$, while for (effective) repulsive
interaction, $u(n)$ grows super-linearly w.r.t $n$. As a result, the $f(n_j)$ turns out to be
%\begin{equation}
%\label{eq:fn2}
%f(n_0)=1, \,f(n_j)=\prod_{j=1}^n\, .
%\end{equation}
\begin{equation}
\label{eq:fn2}
f(n_j) = \left\{
         \begin{array}{rl}
              \prod_{j=1}^n \left[\frac{w_j}{j^\delta}\right]\,,  & \text{if} ~~j \geq 1 \,, \\
              1 \,,                                                 & \text{if} ~~j = 0 \,.
\end{array} \right.
\end{equation}

%%%%%%%%%%%%%%%%%%%%%%%%%%% GCE
\subsection{Grand Canonical Ensemble Formalism}

We now present the general GCE approach that deal with the condensation phase transition in ZRPs~\cite{Evans00}.
The grand canonical partition function is defined by
\begin{equation}
\label{eq:ZL}
Z_L(z)= \sum_{n=0}^{\infty} z^n Z_{L,n} \, .
\end{equation}
where $z$ is the fugacity. Physically it is the average hopping rate. Applying Eq.(\ref{eq:ZLN}) to
Eq.(\ref{eq:ZL}) one obtains
\begin{equation}
\label{eq:ZLz}
Z_L(z)= \sum_{\{n_l=0\}}^{\infty} z^{\sum n}~\prod_{l=1}^L f(n_l) = \prod_{l=1}^L \mathscr{F}_l(z) \, ,
\end{equation}
where $\mathscr{F}_l(z)$ is
\begin{equation}
\label{eq:Fz}
\mathscr{F}_l(z) = \sum_{n=0}^{\infty} z^n~f_l(n)  \, .
\end{equation}

The average occupation number of the $i$-th node $m_i$ reads
\begin{equation}
\label{eq:mi1}
m_i \equiv \langle n_i\rangle = \sum_{\{n_i\}} n_i \mathscr{P}(\{n_i\}) \,.
%= x_i \frac{\partial}{\partial x_i} \ln \mathscr{F} (x_i) \,
\end{equation}
The self-consistent equation $\rho = \sum m/L$ determines the fugacity.

By introducing the effective fugacity $x_i$,
\begin{equation}
\label{eq:xi}
x_i \equiv zw_i = zk_i/\sum k \,,
\end{equation}
and with the help of Eq.-s (\ref{eq:Pn}), (\ref{eq:fn2}), and (\ref{eq:Fz}), $m_i$ can be written as
\begin{equation}
\label{eq:mi2}
m_i = x_i \frac{\partial}{\partial x_i} \ln \mathscr{F} (x_i) \,,
\end{equation}
in which
\begin{equation}
\label{eq:Fx}
\mathscr{F} (x_i) = \sum_{n=0}^\infty\frac{x_i^n}{(n!)^\delta} \,.
\end{equation}

We will apply the above GEC approach to the ZRP on SF networks, and this is the major task in the following
sections.
%We predict the finite density scaling laws for the steady state, as well as the relaxation dynamics.

\section{Finite Density Scaling Law of The Steady State}

\subsection{Analytical Results}

The so-called complete condensation phase transition in ZRPs on SF network with a given density $\rho$ had
been revealed in recent studies~\cite{Noh05a}. The structural inhomogeneity of SF networks plays an important
role. To summarize, when the complete condensation occurs, almost a whole fraction of particles are condensed
onto a few high-degree nodes. As we will see in the following, a variable finite density have some nontrivial
effects on the condensation phase transitions.

Although the inter-particle interactions are determined by the parameter $\delta$ in hopping rate function,
only the case of $0 < \delta < 1$ is what we are concerned in current paper. For when $\delta=0$, the model
is equivalent to the disordered ZRP with $n$-independent hopping rate functions, and had already been studied
in~\cite{Krug00,Evans96,Jain03}. The analytical solution in this case is simply $m_i=x_i/(1-x_i)$.
When $\delta=1$, the hopping rate function becomes $u(n) = n$, which is proportional to the number of particles.
Physically this corresponds to fully independent motion of particles, and the system consists of $N$
non-interacting random particles. One may solve the occupation number distribution $m_i(z)=zk_i$~\cite{Noh05a},
as expected.

We then pay our attention on $0 < \delta < 1$. To solve the average occupation number $m_i$ in this general case,
one has to resort to the approximate expression of $\mathscr{F}(x_i)$ in Eq.~(\ref{eq:Fx}), since in this case
the analytically closed form does not exist. The saddle-point approximations can be borrowed to finish the task
for the series in Eq.(\ref{eq:Fx}). For large $x$, it can be shown~\cite{Noh05a} that
\begin{equation}
\label{eq:sp}
\mathscr{F}(x) \approx x^{(1-\delta)/2\delta} e^{\delta x^{1/\delta}} \,.
\end{equation}
For small $x$, the series in Eq.(\ref{eq:Fx}) can be approximated by summing over a few lowest order terms.
As a result, one obtains $m_i \simeq x_i$ for $x_i\ll 1$, and $m_i \simeq x_i^{1/\delta}$ for
$x_i\gtrsim 1$.
%\begin{equation}
%\label{eq:m_x}
%m_i \simeq \left\{
% \begin{array}{cl}
%   x_i & \mbox{for $x_i\ll 1$}, \\
%   x_i^{1/\delta} & \mbox{for $x_i\gtrsim 1$} \ .
% \end{array}
% \right.
%\end{equation}

The hub owns the most links to another nodes, its fugacity is $z=m_\mathrm{hub}^{\delta} \sum k/k_\mathrm{max}$.
The effective fugacities of the rest of nodes are $x_i \equiv zk_i/\sum k=k_i/k_c$, where $k_c$ is the
\emph{crossover degree}, and defined as
\begin{equation}
\label{eq:kc1}
k_c=k_\mathrm{max}/m_\mathrm{hub}^{\delta} \,.
\end{equation}
The minimum effective fugacity is obviously $x_{min} = k_{min}/k_c$. The $k_c$ is determined by the self-consistency
condition $\rho=\sum m/L$ and the possible form of the minimum effective fugacity $x_{min}$.
%Note that now the density $\rho$ is a variable quantity.
Let us assume at first $x_{min} \gtrsim 1$, then the self-consistency condition reads
\begin{equation}
\label{eq:rho1}
\rho=\overline{k^{1/\delta}}/k_c^{1/\delta} \,.
\end{equation}
The $\overline{(\cdots)}$ in above equation represents the average of $(\cdots)$, $\overline{k^{1/\delta}}$
is just
\begin{equation}
\label{eq:kb}
\overline{k^{1/\delta}} \equiv \sum k^{1/\delta}/L \,,
\end{equation}
by definition. In thermodynamic limit $L \rightarrow \infty$, the summation (over all nodes) in
Eq.(\ref{eq:kb}) can be replaced by integral, hence
%($i=1$, $\ldots$, $L$),
%\textit{i.e.}, the minimum degree $k_\mathrm{min} \gtrsim k_c$,
\begin{equation}
\label{eq:kd}
\overline{k^{1/\delta}} = \int dk ~k^{1/\delta} ~\mathscr{P}_{deg}(k) \,,
\end{equation}
%$\overline{k^{1/\delta}}=\int dk k^{1/\delta} \mathscr{P}_{deg}(k)$.
By means of Eq.(\ref{eq:rho1}), the crossover degree $k_c$ w.r.t. the density can be obtained, i.e.,
\begin{equation}
\label{eq:kc1}
k_c \sim \Big(\overline{k^{1/\delta}}/\rho\Big)^\delta \,, ~~\delta>\delta_c \,.
\end{equation}
%for $\delta>\delta_c$.
Note that $k_c$ is density dependent and scales as $k_c \sim \rho^{-\delta}$.
%The non-divergent condition of $\overline{k^{1/\delta}}$ leads to
%$\delta > \delta_c = 1/(\gamma-1)$. In such case,
The average occupation number $m_k$ on a node with degree $k>k_c$ for steady state becomes
\begin{equation}
\label{eq:mk1}
m_{k>k_c}=\rho k^{1/\delta}/\overline{k^{1/\delta}} \,.
\end{equation}
On the hub, it scales as
\begin{equation}
\label{eq:mhub}
m_\mathrm{hub} \sim \rho k_\mathrm{max}^{1/\delta}/\overline{k^{1/\delta}}
\sim \rho L^{\delta_c/\delta} \,,
\end{equation}
where the last relationship comes from the fact that
$k_\mathrm{max} \approx k_{min} L^{1/(\gamma-1)} \sim L^{\delta_c}$
for the SF network. Since $\delta_c/\delta < 1$, the sub-linear increasing of $m_\mathrm{hub}$ on $L$
suggests that no condensation transition occurs, the whole system is in a single non-condensation phase.
Physically, this is due to the existence of the lower bound degree $k_\mathrm{min}$. When
$k_c<k_\mathrm{min}$, the condensation transition is fully suppressed and only a single \emph{fluid-like}
phase exists for $\delta > \delta_c$.

On the other hand, if we assume $x_{min}\ll 1$, or equivalently $k_c \gg k_\mathrm{min}$, the density can
be decomposed into a fluid part $\rho_f$ and a condensed part $\rho_c$:
\begin{equation}
\rho = \rho_{f}+\rho_{c} \,,
\end{equation}
with
\begin{equation}
\rho_f=L^{-1} \sum_{k_i<k_c}x_i \,\,\,, \rho_c=L^{-1}\sum_{k_i>k_c}x_i^{1/\delta} \,.
\end{equation}
In the thermodynamic limit $L \rightarrow \infty$, the summands in $\rho_c$ and $\rho_f$ read respectively,
\begin{eqnarray}
\label{eq:rho2}
\rho_f &=& k_c^{-1} \int_{k_\mathrm{min}}^{k_c} dk ~k~\mathscr{P}_{deg}(k) \,, \\
\rho_c &=& k_c^{-1/\delta}\int_{k_c}^{k_\mathrm{max}} dk ~k^{1/\delta}~\mathscr{P}_{deg}(k) \, .
\end{eqnarray}
The integral $\int_{k_\mathrm{min}}^{k_c} dk ~k ~\mathscr{P}_{deg}(k)$ is finite such that a condensation
phase transition may occur if $\rho_{f}$ vanishes as $\sim k_c^{-1}$. As a result,
\begin{equation}
\label{eq:rho_c}
\rho_c \rightarrow \rho=k_c^{-1/\delta}\int_{k_c}^{k_\mathrm{max}}dk ~k^{1/\delta-\gamma} \, .
\end{equation}
The self-consistency condition requires the integral in Eq.~(\ref{eq:rho_c}) divergent as $L \rightarrow \infty$.
This is to require the exponent part in above integral, $1/\delta-\gamma > -1$, such that $\delta \le \delta_c$.
One sees that $k_c$ is the degree at which the fluid-like phase crossovers to the condensation phase, and scales
as $k_c \sim (\ln k_\mathrm{max}/\rho)^{\delta_c}$ for $\delta=\delta_c$. For $\delta < \delta_c$, it scales as
\begin{equation}
\label{eq:kc2}
k_c \sim k_\mathrm{max}^{1-\delta/\delta_c}/\rho^\delta, ~~\delta < \delta_c \, .
\end{equation}
%This shows a scaling behavior of $k_c$ w.r.t. the density, comparing
%to that in Eq.~(\ref{eq:kc1}) for $\delta >\delta_c$.
For an SF network with finite size $L$, Eq.~(\ref{eq:kc2}) is reduced to $k_c \sim L^{\delta_c-\delta}/\rho^\delta$
for $\delta <\delta_c$. Note that the presence of the factor $\rho^{-\delta}$ implies a decreasing of $k_c$ as the
density increases. Further, the average occupation number scales as
$m_k \sim \rho^{\delta_c}k/\left( \ln k_{\mathrm{max}}\right)^{\delta_c}$ ($k<k_c$),
and $m_k \sim \rho k^{1/\delta_c}/\ln k_{\mathrm{max}}$ ($k>k_c$) for $\delta=\delta_c$.

For $\delta<\delta_c$, we have
\begin{equation}
\label{eq:mk21}
m_k \sim \left\{
 \begin{array}{rl}
   \rho^\delta k/k^{1-\delta/\delta_c}_{\mathrm{max}} , & \mbox{$k<k_c$}, \\
   \rho k^{1/\delta}/k^{1/\delta-1/\delta_c}_{\mathrm{max}} , & \mbox{$k>k_c$} \ ,
 \end{array}
 \right.
\end{equation}
%$m_k \sim k_\mathrm{max}^{1-\delta/\delta_c}/\rho^\delta$ for
%$\delta <\delta_c$.
For a finite number of nodes $L$, Eq.~(\ref{eq:mk21}) becomes
\begin{equation}
\label{eq:mk2}
m_k \sim \left\{
 \begin{array}{rl}
   \rho^\delta k/L^{\delta_c-\delta} \,, & \mbox{$k<k_c$} \,, \\
   \rho k^{1/\delta}/L^{\delta_c/\delta-1} \,, & \mbox{$k>k_c$} \,,
 \end{array}
 \right.
\end{equation}
for the fluid-like phase ($k<k_c$) and the condensation phase ($k>k_c$), respectively. In contrast to the
case with a fixed density, now the density $\rho$ comes in and has effects on the average occupation number
and hence the condensation phase transition.

We summarize our predicted results in Table~\ref{table1} for readability. The finite density scaling relations
are explicitly presented.
\begin{table}[t]
\caption{The scaling relations of the crossover degree $k_c$ and the average degree occupation number $m_k$
w.r.t. the finite density $\rho$ and the parameter $\delta$.
\label{table1}}
%\centering
\begin{ruledtabular}
\begin{tabular}{cccc}
$\delta$ & $m_{k < k_c}$ & $k_c$ & $m_{k > k_c}$  \\
\hline
$\delta > \delta_c$ & --- & $\Big(\overline{k^{1/\delta}}/\rho\Big)^\delta$
& $\rho k^{1/\delta}/\overline{k^{1/\delta}}$  \\
$\delta = \delta_c$ & $\rho^{\delta_c}k/\left( \ln k_{\mathrm{max}}\right)^{\delta_c}$
& $\left( \ln k_{\mathrm{max}}/\rho \right)^{\delta_c}$
& $\rho k^{1/\delta_c}/\ln k_{\mathrm{max}}$ \\
$\delta < \delta_c$ & $\rho^\delta k/k^{1-\delta/\delta_c}_{\mathrm{max}}$
&  $k^{1-\delta/\delta_c}_{\mathrm{max}}/\rho^\delta$
& $\rho k^{1/\delta}/k^{1/\delta-1/\delta_c}_{\mathrm{max}}$ \\
\end{tabular}
\end{ruledtabular}
\end{table}

%\begin{table}[t]
%\caption{The scaling relations of the crossover degree $k_c$, the average
%degree occupation number $m_k$, and the average occupation number of hub $m_{hub}$
%w.r.t the number density $\rho$.
%\label{table1}}
%\centering
%\begin{ruledtabular}
%\begin{tabular}{ccccc}
%$\delta$ & $k_c$ & $m_{k < k_c}$ & $m_{k > k_c}$ & $m_{\mathrm{hub}}$ \\
%\hline
%$\delta > \delta_c$ & $\overline{k}/\rho^\delta$ & ---
%& $\rho k^{1/\delta}/\overline{k}^{1/\delta}$ & $\mathscr{O}(\rho L^{\delta_c/\delta})$ \\
%$\delta = \delta_c$ & $\left( \ln k_{\mathrm{max}}/\rho \right)^{\delta_c}$
%& $\rho^{\delta_c}k/\left( \ln k_{\mathrm{max}}\right) ^{\delta_c}$
%& $\rho k^{1/\delta_c}/\left( \ln k_{\mathrm{max}}\right) $ & $\mathscr{O}\left(\rho L/\ln L\right) $ \\
%$\delta < \delta_c$ & $k^{1-\delta/\delta_c}_{\mathrm{max}}/\rho^\delta$
%& $\rho^\delta k/k^{1-\delta/\delta_c}_{\mathrm{max}}$
%& $\rho k^{1/\delta}/k^{1/\delta-1/\delta_c}_{\mathrm{max}}$ & $\mathscr{O}\left(\rho L\right) $ \\
%$\delta = 0$ & $k_{\mathrm{max}}$ & $k/\left( k_{\mathrm{max}}-k\right)$ & --- & $\rho L$ \\
%\end{tabular}
%\end{ruledtabular}
%\end{table}

%%%%%%%%%%%%%%%%%%%%%%%%%%%%

\subsection{Numerical Simulations for Condensation Transition at Steady State}

In order to verify the theoretical predictions of the scaling laws by previous GCE analysis, we perform
the following Monte Carlo (MC) simulations.
The underlying SF network is generated using the Bar\'abasi-Albert model~\cite{BA99}, with a degree
distribution $\mathscr{P}(k) \sim k^{-\gamma}$ with $\gamma=3$, and hence leads to a critical value
$\delta_c=0.5$. The initial conditions for the simulations are random distribution.
%After $\sim 10^{15}$ jumps the quantities of interest had apparently converged and
%suggested the equilibration of the dynamics or
After the realization of the steady state, we keep the system running many times and then measure the
quantities interested.

\begin{figure}[t!]
\includegraphics[width=0.42\textwidth]{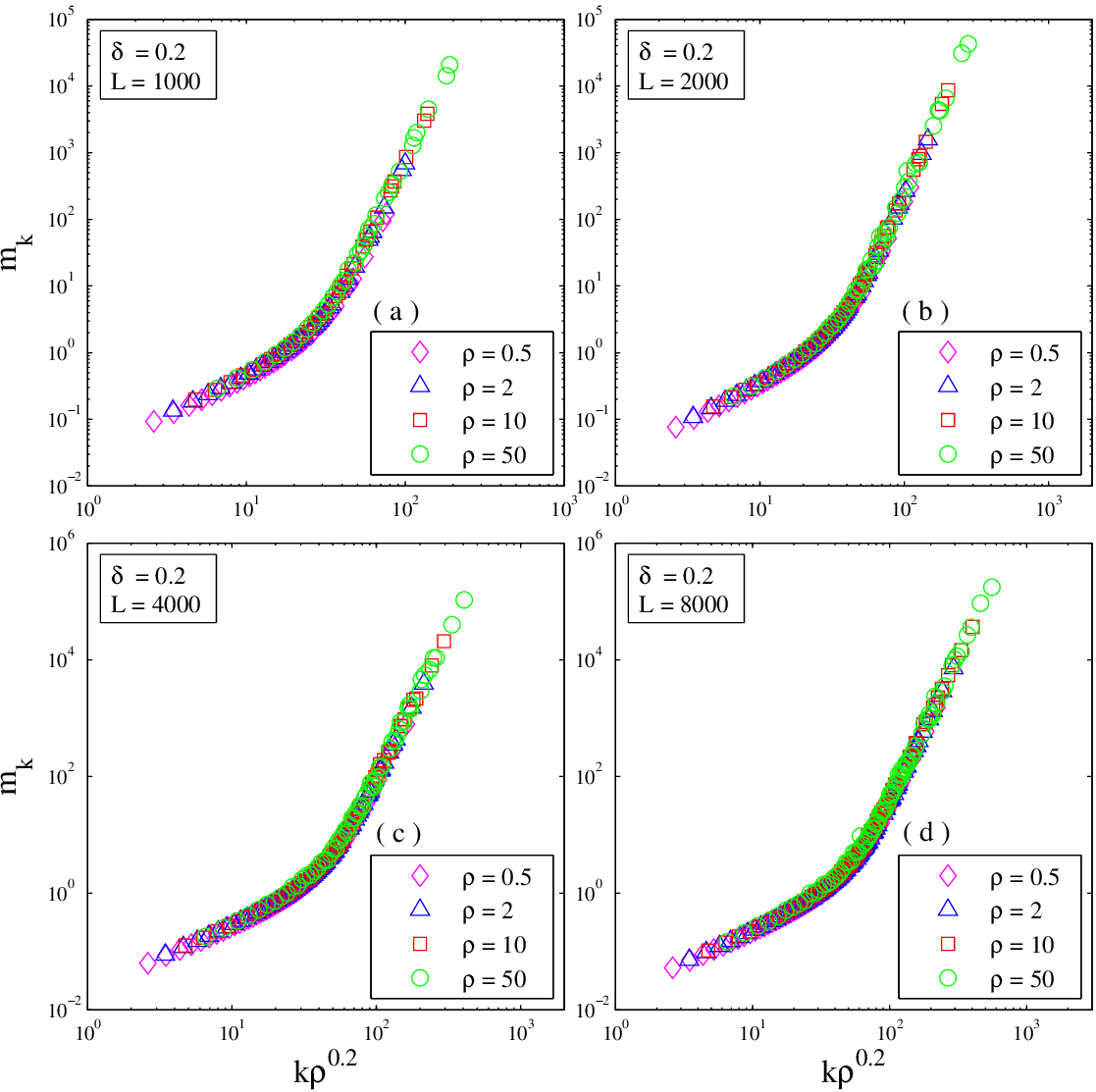}
\centering
\caption{(Color online) The scaling relations of the average occupation number $m_k$ versus $k\rho^\delta$ at
$\delta=0.2$ for different SF network sizes: (a) $L=1000$, (b) $L=2000$, (c) $L=4000$, (d) $L=8000$, at various
densities $\rho=0.5$ (purple open diamonds), $\rho=2$ (blue open triangles), $\rho=10$ (red open squares),
$\rho=50$ (green open circles), respectively.\label{fig1}}
\end{figure}

%We have checked the finite size scaling at $\delta=0.2$ and at various densities.
%$\rho=0.5$, $2$, $10$, and $50$.
%At $\rho=2$, the results in \cite{Noh05a} is recovered. In addition, as expected from
%Eq.~(\ref{eq:kc2}), the crossover degree $k_c$ decreases when $\rho$ increases.
To see the presence of the finite density effect, in Fig.~\ref{fig1}, the scaling relations of $m_k$ versus
$k\rho^\delta$ are shown at $\delta=0.2 < \delta_c$ for different network sizes $L=1000$, $2000$, $4000$, and $8000$,
and various densities, $\rho=0.5$ (purple open diamonds), $\rho=2$ (blue open triangles), $\rho=5$ (red open
squares), $\rho=10$ (green open circles), respectively. As expected from Eq.(\ref{eq:mk2}), for $k<k_c$ (i.e.,
on the left hand side of the crossover degree $k_c$), it is obvious that $m_k \sim k\rho^\delta L^{\delta-\delta_c}$,
hence for given network with size $L$, $m_k \propto k\rho^\delta$ for a fixed $\delta$; while for $k>k_c$
(i.e. the right hand side of the crossover degree $k_c$), $m_k \sim k^{1/\delta} \rho L^{1-\delta_c/\delta}$,
for a given network with size $L$, $m_k \sim k^{1/\delta} \rho \propto (k\rho^\delta)^{1/\delta}$. In other
words, in both cases, the average occupation number is an scaling function of $k\rho^{\delta}$ at given $L$,
namely, $m_k \sim \mathscr{G}(k\rho^\delta)$, where $\mathscr{G}$ is some scaling function. One finds in
Fig.~\ref{fig1} the simulation data nicely collapse onto the same curve, hence fully support our theoretical
predictions.
%and the crossover degree $k_c$ decreases when the density increases.

\begin{figure}[t!]
\includegraphics[width=0.42\textwidth]{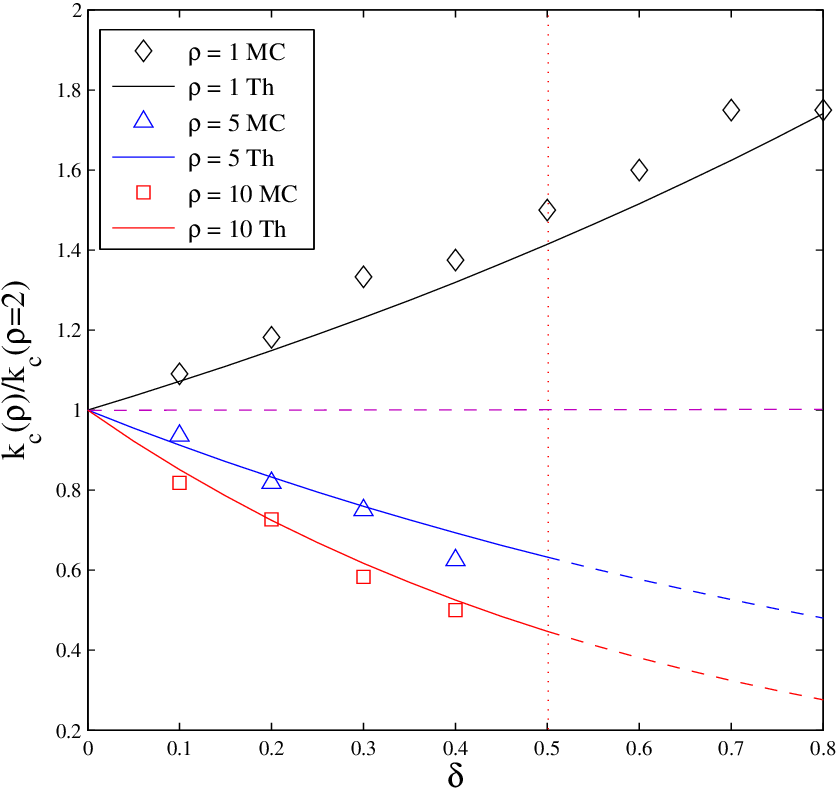}
\centering
\caption{(Color online) The presence of the finite density effect in the crossover degree ratios
$k_c(\rho)/k_c(\rho=2)$ versus the parameter $\delta$ with an SF network size $L=4000$, for various densities
$\rho=1$ (black open diamonds and solid line), $5$ (blue open triangles and solid line), and $10$ (red open
squares and solid line), respectively. The solid lines correspond to the theoretical calculation and the symbols
to the MC simulation data. Note that $k_c$'s do not exist in regime $\delta>\delta_c$ ($\delta_c=0.5$ is the
vertical red dotted dash line) for $\rho=5$ and $\rho=10$ due to the lower bound of degree $k_{\mathrm{min}}$,
two theoretical dashed lines are only formally drawn in that region.\label{fig2}}
\end{figure}

The predicted finite density effects can be viewed from the other perspective.
%this can be seen clearly from Fig.~\ref{fig1}.
From Eq.-s~(\ref{eq:kc1}) and (\ref{eq:kc2}), a $\rho$-dependent ratio can be derived, i.e.,
$k_c(\rho_2)/k_c(\rho_1)=(\rho_1/\rho_2)^\delta$. This ratio eliminates the effect of the proportional constants
in $k_c$. We have to emphasize at this point that, without taking into consideration of the variable density,
this ratio should be a density \emph{independent} constant, i.e., $k_c(\rho_2)/k_c(\rho_1)=1$.
In Fig.~\ref{fig2} we use $k_c$ at $\rho_1=2$ as a benchmark, plotted theoretical ratios $k_c(\rho)/k_c(\rho=2)$
with various densities, $\rho=1$ (black solid line), $\rho=5$ (blue solid line), and $\rho=10$ (red solid line),
respectively, as a function of $\delta$ at given network size $L=4000$. The purple dashed horizontal line shown
in Fig.~\ref{fig2} just corresponds to the density independent ratio, which is unity obviously.
However, MC simulations show the density dependent ratios for $\rho=1$ (open black diamonds), $\rho=5$ (open blue
triangles), and $\rho=10$ (open red squares), respectively, and agree with our theoretical predictions very well.
In principle, $k_c$'s in $\delta> \delta_c$ regime do not exist due to the existence of lower bound $k_{\mathrm{min}}$
for densities $\rho=5$ and $\rho=10$, hence we formally draw the theoretical results by blue ($\rho=5$) and red
($\rho=10$) dashed lines.

%\begin{figure}
%\includegraphics[width=\columnwidth]{fig1}\\
%\caption{(Color online) The finite density effects and the the finite size scaling
%relations between the average occupation number $m_k$ and different sizes of SF
%network for $\delta=0.2$ and densities: (a) $\rho=0.5$, (b) $\rho=2$, (c) $\rho=10$,
%(d) $\rho=50$, respectively. In each panel, the $m_k$ vs. $k/L^{1-\delta}$
%scaling is drawn for $L = 1000$ (purple open diamonds), $L = 2000$ (blue open
%triangles), $L = 4000$ (red open squares), $L = 8000$ (green open circles),
%respectively. The crossover degree $k_c$ (the vertical red dash line in each panel
%corresponds to $L=8000$) between the fluid-like phase and condensation phase
%decreases as the density $\rho$ increases.
%\label{fig1}}
%\end{figure}

%%%%%%%%%%%%%%%%%%%%%%%%%%%%%%%% relax. dyn.
\section{Finite Density Scaling Law of Relaxation Dynamics}

%We now turn to the relaxation dynamics in the condensation phase.
The preceding theoretical predictions show that the occupation number distribution in the steady state is
determined not only by the degree distribution $\mathscr{P}_{deg}(k)$ being independent of any other
characteristics of underlying networks, but also the density of the particles hopping on the networks.
However, the relaxation dynamics of ZRP is getting more complicated due to the effects of both the structure
of underlying networks and the density.

So far only has a little been known about the dynamics of condensation in ZRP~\cite{Godre03,Gross03,Godre17}.
The finite density effects on the relaxation dynamics is hence one of the important issues to be addressed.
In the following we only consider the case of $\delta <\delta_c$, in which a condensation phase transition
occurs.we rely on the MC simulations and the scaling ansatz in order to understand the dynamical
scaling properties.

Scaling ansatz is helpful for obtaining the qualitative features of relaxation dynamics. Previous
studies~\cite{Godre03,Gross03} revealed that, particles form macroscopic condensate by successive coarsening
processes of the small clusters.
The smaller clusters merge into larger ones, and grow until a single macroscopic condensate forms. The scaling
hypothesis leads to the power-law growth of the relaxation time with the system size as $t_R \sim L^\beta$,
where $\beta=1-\delta$ is the dynamic exponent.

The highly heterogeneous structure of SF networks deeply affect the relaxation dynamics of interacting
particles system defined on them. It was hypothesized in \cite{Noh05a} that the relaxation process follows
some hierarchical dynamics. There exist two characteristic time-dependent degree scales $k_{\Omega c}(t)$ and
$k_{\Omega}(t)$, which plays the role of the crossover degree $k_c$, and the role of $k_{max}$, respectively.
During transient  time $t$, a subnetwork of small degree nodes is equilibrated first and the equilibrated
subnetwork keep expanding until $k_{\Omega}(t)$ grows in time and eventually reaches $k_{max}$.

It is natural to assume that a similar hierarchical dynamics occurs in a variable finite density situation.
Starting from a fully random distribution at initial time, all particles behave as random walkers without
interaction in a short period of time.
%and has a $m_k \sim k$ behavior, up to some density-dependent factor.
Later on the distribution of particles further evolves and eventually approaches the steady state.
Suppose there exist two characteristic degree scales in hierarchical relaxation dynamics.
Those nodes with degree $k \leq k_\Omega$ consist of a smaller equilibrated subnetwork, denoted below by,
say, $\Omega$, of size $L_\Omega < L$ in a transient time $t \ll t_R$, where $t_R$ is the relaxation time.
Inside this subnetwork, the other characteristic degree $k_{\Omega\mathrm{c}}$ plays the role of crossover
degree $k_c$ of the whole network. The equilibrated subnetwork keeps growing to proceed to the higher
hierarchy until $k_\Omega$ reaches $k_{\mathrm{max}}$ of the network, the steady state of the ZRP is then
formed. We may refer to those nodes with $k_{\Omega}$ as the \emph{temporary hub}~\cite{Noh05b}. The average
occupation number $m_{k_\Omega}$ versus the transient time $t$, according to the hierarchical dynamics,
scales as
\begin{equation}
m_{k_\Omega} \sim \rho k_\Omega^{1/\delta_c} \sim t^{1/\nu} \,,
\end{equation}
with $\nu=1-\delta$ being the dynamic exponent~\cite{Noh05b}. Clearly $k_\Omega$ evolves with time, however,
we emphasize at this moment that $k_\Omega$ is also density dependent, i.e.,
\begin{equation}
\label{eq:kot}
k_\Omega \sim t^{\delta_c/\nu}/\rho^{\delta_c}  \, .
\end{equation}
We now turn to the other scale in the transient state: the crossover degree $k_{\Omega c}$. Just like $k_c$'s
scaling in the steady state of the network, it scales as
\begin{equation}
\label{eq:koc}
k_{\Omega c} \sim k_\Omega^{1-\delta/\delta_c}/\rho^\delta \,,
\end{equation}
and again, a density dependent factor $\rho^{-\delta}$ comes in.
%for $\delta <\delta_c$ scales as $\delta=\delta_c$ scales as
%$k_{\Omega c} \sim (\ln k_\Omega/\rho)^{\delta_c}$,
%k_{\Omega c} \sim k_\Omega^{1-\delta/\delta_c}/\rho^\delta  \ .

The average occupation number of a node within the equilibrated network $m_k$ reads
\begin{equation}
\label{eq:mko}
m_k \sim \left\{
\begin{array}{rl}
\rho^\delta k/k^{1-\delta/\delta_c}_{\Omega}, & \mbox{$k<k_{\Omega\mathrm{c}}$} \,, \\
\rho k^{1/\delta}/k^{1/\delta-1/\delta_c}_{\Omega} , & \mbox{$k_{\Omega\mathrm{c}}<k<k_{\Omega}$} \, .
\end{array}
\right.
\end{equation}
As a result, by applying Eq.~(\ref{eq:kot}), the scaling law of time- and density-dependent $m_k$ can be
obtained as
\begin{equation}
\label{eq:mko}
m_k \sim \left\{
\begin{array}{rl}
k\rho^{\delta_c}/t^{(\delta_c-\delta)/\nu} \,, & \mbox{$k<k_{\Omega\mathrm{c}}$} \,, \\
(k\rho^{\delta_c})^{1/\delta}/t^{(\delta_c-\delta)/\delta \nu} \,, & \mbox{$k_{\Omega\mathrm{c}}<k<k_{\Omega}$} \, .
\end{array}
\right.
\end{equation}
These explicit scaling laws of $m_k(t)$ versus $k\rho^{\delta_c}$ for some transient time $t$ can be unified
in a single relation $m_k \sim \mathscr{H}(k\rho^{\delta_c})$, in which $\mathscr{H}$ is some scaling function,
for given $t$ and $\delta$. MC simulations have to be performed to verify the above analytical predictions.

%\begin{table}[t]
%\caption{The scaling relations of the crossover degree $k_{\Omega c}$ and
%the average degree occupation number $m_k$ w.r.t. the finite density $\rho$
%and the parameter $\delta$.
%\label{table2}}
%\centering
%\begin{ruledtabular}
%\begin{tabular}{ccc}
%$m_{k <k_{\Omega\mathrm{c}}}$ & $k_{\Omega\mathrm{c}}$ & $m_{k>k_{\Omega\mathrm{c}}}$  \\
%\hline
%$\rho^\delta k/k^{1-\delta/\delta_c}_{\Omega}$ & $k^{1-\delta/\delta_c}_{\Omega}/\rho^\delta$
%& $\rho k^{1/\delta}/k^{1/\delta-1/\delta_c}_{\Omega}$
%\end{tabular}
%\end{ruledtabular}
%\end{table}

\begin{figure}[t!]
\includegraphics[width=0.42\textwidth]{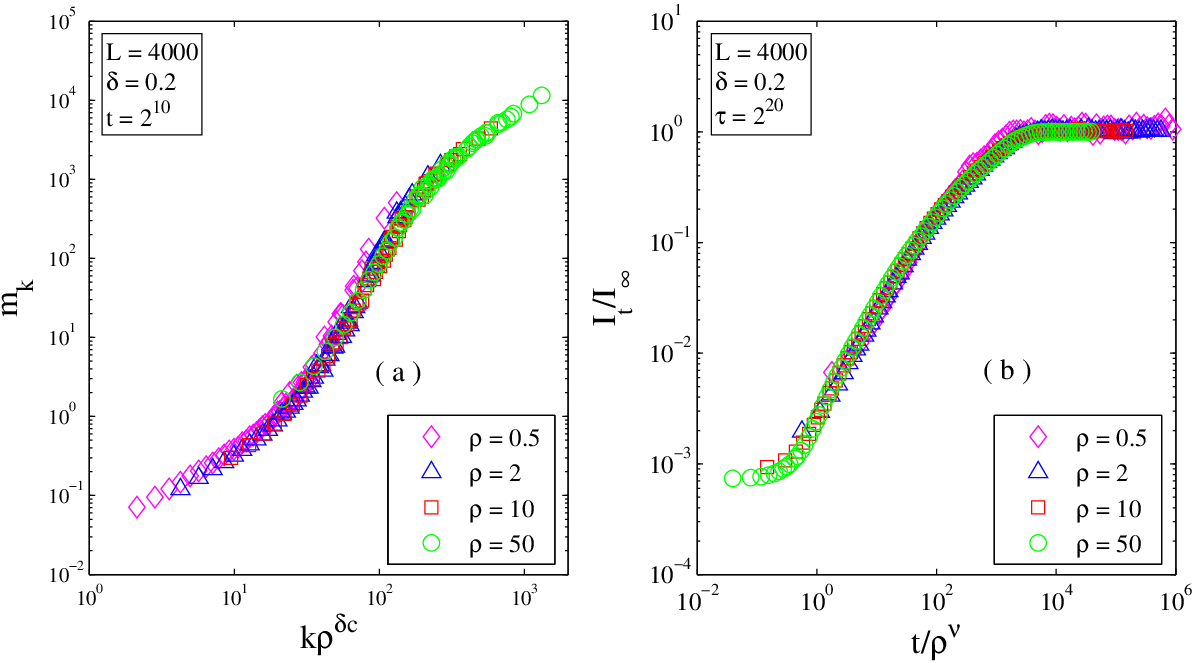}
\centering
\caption{(Color online) Data collapse plot of (a) the average occupation number $m_k$ versus $k\rho^{\delta_c}$
at the transient time $t=2^{10}$, (b) the evolution of ratio of IPR to its steady state limit, $I_t/I_{\infty}$,
versus $t/\rho^\nu$, with $\nu={1-\delta}$ being the dynamic exponent in the relaxation dynamics, at
$\delta = 0.2$ for network size $L=4000$, and various densities $\rho=0.5$ (purple open diamonds), $\rho=2$
(blue open triangles), $\rho=10$ (red open squares), and $\rho=50$ (green open circles), respectively.
%In panel (b), a total evolution time $\tau=2^{20}$is taken in the simulations.
\label{fig3}}
\end{figure}

The results of MC simulations on the finite density effects of the relaxation dynamics are shown in
Fig.~\ref{fig3}(a).
Note the different time scales involved in our relaxation dynamics simulations: the transient time scale $t$,
the relaxation time scale $t_R$, the evolution time scale of the system $\tau$, and $t \ll t_R \ll \tau$.
%where is in our simulation.
We took the evolution time $\tau=2^{20}$, which is long enough for system to evolve towards the steady state.
The average occupation number $m_k(t)$ at transient time $t=2^{10}$, versus $k\rho^{\delta_c}$ with a given
network size $L=4000$, but with various densities $\rho=0.5$ (purple open diamonds), $\rho=2$ (blue open
triangles), $\rho=10$ (red open squares), and $\rho=50$ (green open circles), respectively, are shown.

As one can see from the left panel in the figure, the collapse of simulation data of $m_k$ w.r.t.
$k\rho^{\delta_c}$ indicates the scaling law in the evolution of the relaxation for different densities.
To understand the condensation dynamics, an appropriate characteristic order parameter is the infinity time
limit of inverse participation ratio (IPR) $I_t$~\cite{Noh05a}, which is defined as
\begin{equation}
\label{eq:It}
I_{\infty} \equiv \lim_{t \rightarrow \infty} I_t = \lim_{t \rightarrow \infty} \frac{1}{N^2} \sum_{i=1}^{L} n^2_i(t) \,.
\end{equation}
According to the hierarchical dynamics ansatz in condensation processes, the occupation number of particles on
the temporary hub dominates in an equilibrated sub-network, note that the degree of the temporary hub $k_{\Omega}$
scales as $t^{\delta_c/\nu}/\rho^{\delta_c}$ as in Eq.~(\ref{eq:kot}), combining with
$L_\Omega\sim k_\Omega^{1/\delta_c}$, the size of the equilibrated sub-network satisfies $L_\Omega\sim t/\rho^\nu$
for a given transient time $t$. Hence the IPR ratio, describing the extent of condensation for a given network size
$L$, possesses the same scaling relation $I_t/I_{\infty} \sim t/\rho^\nu$, while for the long run $\tau \gg t_R$ the
steady state is reached and the particles on the hub dominate.
%$\tau \sim \rho^\nu k_{max}^{\nu/\delta_c} \sim \rho^\nu L^\nu$, one then obtains.

%%%%%%%%%%%%%%%   updated 2019.10
%the temporary hub dominates in an equilibrated sub-network, while for the long run $\tau \gg t_R$ the steady
%state is reached and the particles on the hub dominate, hence the IPR ratio possesses a scaling relation
%$I_t/I_{\infty} \sim t/\rho^\nu$ for a given network size $L$, if one combines $L_\Omega\sim k_\Omega^{1/\delta_c}$
%and Eq.~(\ref{eq:kot}), that is $k_\Omega \sim t^{\delta_c/\nu}/\rho^{\delta_c}$.
%$\tau \sim \rho^\nu k_{max}^{\nu/\delta_c} \sim \rho^\nu L^\nu$, one then obtains.

In Fig.~\ref{fig3}(b), we plot the scaling relations of $I_t/I_{\infty}$ versus $t/\rho^\nu$ at $\delta=0.2$
for $\rho=0.5$, $\rho=2$, $\rho=10$, and $\rho=50$, respectively. A total $\tau=2^{20}$ time-step evolution
has been fulfilled to ensure the realization of the steady state, i.e., $I_{\tau} \rightarrow I_{\infty}$.
One immediately sees that the increasing density naturally postpones the occurrence of the condensation transition,
and the collapse of data of different densities nicely confirms our prediction.

%\begin{figure}
%\includegraphics[width=\columnwidth]{fig6}\\
%\caption{(Color online) Data collapse plot of the occupation distribution $m_k$ vs.
%$k\rho^{\delta_c}$ at transient time (a) $t=2^8$, (b) $t=2^{10}$, respectively,
%for network size $L=4000$ and $\delta = 0.2$. In each panel, data at
%various densities $\rho=0.5$ (purple open diamonds), $\rho=2$ (blue open
%triangles), $\rho=10$ (red open squares), $\rho=50$ (green open circles)
%collapse.
%\label{fig6}}
%\end{figure}

\section{Summary}\label{sec:5}

In summary, we explored the finite density effects on both the steady and dynamical properties of the ZRP
on SF network with hopping rate function $u(n)=n^{\delta}$. By means of the GCE method, we discovered that
the steady state condensation phase transition is driven not only by the disorder of the underlying SF
network, but also by the finite density of the interacting particles. The crossover degree $k_c$ and the
average occupation number $m_k$ were found to be both density dependent, and satisfy the corresponding
scaling laws. In contrast to the ZRP on SF network with fixed density, the ratios of the crossover degree
$k_c$ for two different densities at given size of network exhibit a non-constant behavior.

The influences of the density on the relaxation dynamics was also analytically investigated.
The hierarchical characteristics of the relaxation dynamics renders the process proceeding from nodes with
lower degrees toward those with higher degrees. With the help of the scaling ansatz, the scaling laws of
the condensation transition (to the steady state) were predicted.
At a specific time $t \ll t_R$ the average occupation number scales as $m_k \sim k\rho^{\delta_c}$ for
various densities. The evolution of the ratio of IPR to its steady state limit follows a scaling law
$I_t/I_{\infty} \sim t/\rho^{\nu}$ with dynamical exponent $\nu=1-\delta$.
%It is characterized by the power-law growth of the degree scale
%with a dynamical exponents.
To verify our analytically derived scaling laws, we have performed the Monte Carlo simulations for varied
densities. The corresponding scaling relations are validated very well by numerical experiments.

\acknowledgements
The authors thank the National Natural Science Foundation of China (NSFC) under Grant No. 11505115 for
financial support.

%%%%%%%%%%%%%%%%%%%%%%%%%%%%%%%%%%%%%%%%%%%%%%%%%%%%%%%%%%%%%%%%%%%%%%%%%

\end{document}